\documentclass[a4paper]{jpconf}

\usepackage{amsmath,slashed,booktabs,amssymb}
\usepackage{color}
\usepackage{breakurl}
\usepackage{xspace}
\usepackage{cite}

\usepackage[utf8]{inputenc}

\newcommand{\Br}{\text{Br}}

\newcommand{\mk}{m_K}

\renewcommand{\Im}{\text{Im}\,}

\usepackage[colorlinks]{hyperref}
\def\hhref#1{\href{http://arxiv.org/abs/#1}{arXiv:#1}} 

\begin{document}
\title{%
Probing lepton flavour (universality) violation \\ 
at NA62 and future kaon experiments
}

\author{Lewis~C.~Tunstall,\footnote[0]{Speaker}$^{, 1}$ Andreas Crivellin,$^2$ \\ 
Giancarlo D'Ambrosio$^3$ and Martin Hoferichter$^4$}

\address{$^1${Albert Einstein Centre for Fundamental Physics, 
Institute for Theoretical Physics, \\
University of Bern, Sidlerstrasse 5, CH--3012 Bern, Switzerland} \\
$^2${Paul Scherrer Institut, CH--5232 Villigen PSI, Switzerland} \\
$^3${INFN-Sezione di Napoli, Via Cintia, I--80126 Napoli, Italy} \\
$^4${Institute for Nuclear Theory, University of Washington, Seattle, WA 98195-1550, USA}}

\ead{tunstall@itp.unibe.ch}

\begin{abstract}
Recent results from the LHC's first run have revealed intriguing departures from lepton flavour universality in the semi-leptonic decays of $B$-mesons.  We discuss the complementary role that rare kaon decays can provide in testing new physics explanations of these flavour anomalies. In the framework of minimal flavour violation, we relate the chiral low-energy constants involved in $K\to\pi\ell\ell'$ and $K\to\ell\ell'$ ($\ell = \mu \mbox{ or } e$) with the new physics Wilson coefficients of the $b\to s$ effective Hamiltonian. We comment on the determination of these low-energy constants at NA62 and future kaon experiments, as well as the required improvements in sensitivity necessary to test the $B$-physics anomalies in the kaon sector.
\end{abstract}

\section{Introduction}
With the completion of Run 1 at the LHC, we find ourselves with several indirect hints that physics beyond the Standard Model (SM) may be lurking in the semi-leptonic decays of $B$-mesons.  Among the recent experimental anomalies, two at LHCb have received  considerable attention:
\begin{itemize}
  \item a $2.6\sigma$ signal of lepton flavour universality violation (LFUV) in the measured~\cite{Aaij:2014ora} branching fractions of $B\to K\ell^+\ell^-$ decays $(\ell = \mu \mbox{ or } e)$;
  \item the measured value~\cite{Aaij:2013qta,LHCb:2015dla} of the angular observable $P_5'$~\cite{Descotes-Genon:2013vna} in the decay $B \to K^*\mu^+\mu^-$ deviates from the SM prediction at the $2$--$3\sigma$ level~\cite{Descotes-Genon:2014uoa,Altmannshofer:2014rta,Jager:2014rwa}.
\end{itemize}
Taken separately, each deviation is at most a $3\sigma$ effect; however, when combined with other $b\to s$ transitions, global fits~\cite{Altmannshofer:2015sma,Descotes-Genon:2015uva} indicate that (a) new physics (NP) is preferred over the SM by $4$--$5\sigma$, and (b) the effect is in the $\mu\mu$ modes only.  Expressed in terms of the effective Hamiltonian for $b\to s$ transitions~\cite{Descotes-Genon:2015uva}
\begin{equation}
\label{H_B}
  {\cal H}_\text{eff}^{|\Delta B|=1} = - \frac{4G_F}{\sqrt{2}} V_{tb}V_{ts}^* \sum_{i} C_i^B(\mu) Q_i^B(\mu)+ \text{h.c.}\,,
\end{equation}
the potential NP can be interpreted as contributions to the Wilson coefficients $C_{9,10}^B$ of the semi-leptonic operators 
\begin{align}
\label{Qi_B}
 Q_{9}^B =   \frac{e^2}{32\pi^2}\left[\bar s \gamma^\mu (1 - \gamma_5) b \right]
\sum_{\ell=e,\mu} \left[\bar{\ell} \gamma_\mu  \ell \right] \quad \mbox{and} \quad
Q_{10}^B  = \frac{e^2}{32\pi^2}\left[\bar s \gamma^\mu (1 - \gamma_5) b \right]
\sum_{\ell=e,\mu}  \left[\bar{\ell} \gamma_\mu \gamma_5 \ell \right]\,.
\end{align}

Here we discuss the complementary role that rare kaon decays can provide in testing NP explanations of the $B$-physics anomalies.  Our analysis~\cite{Crivellin:2016vjc} is based on the observation that at low energy scales $\mu \ll m_{t,b,c}$, the strangeness-changing transitions are described in terms of the effective Lagrangian~\cite{Cirigliano:2011ny}
\begin{equation}
\label{Lagr_K}
  {\cal L}_\text{eff}^{|\Delta S|=1} = 
  - \frac{G_F}{\sqrt{2}} V_{ud}V_{us}^* \sum_{i} C_i(\mu) Q_i(\mu) + \text{h.c.} \,,
\end{equation}
which contains semi-leptonic operators
\begin{align}\label{eq:Q11-12}
Q_{7V} =   \left[\bar s \gamma^\mu (1 - \gamma_5) d \right]
\sum_{\ell=e,\mu} \left[\bar{\ell} \gamma_\mu  \ell \right] \quad \mbox{and} \quad
Q_{7A} = \left[\bar s \gamma^\mu (1 - \gamma_5) d \right]
\sum_{\ell=e,\mu}  \left[\bar{\ell} \gamma_\mu \gamma_5 \ell \right]\,, 
\end{align}
that are the $s\to d$ analogues of the $b\to s$ operators $Q_{9,10}^B$ in (\ref{Qi_B}).  In the framework of minimal flavour violation (MFV), the Wilson coefficients of the two sectors \emph{are correlated}, and we use this feature to convert knowledge of $C_{7V,7A}$ into bounds on $C_{9,10}^B$.  The quality of the bounds is hampered by non-perturbative effects from QCD,  which are parametrised in terms of the low-energy constants (LECs) arising in the 3-flavour chiral expansion.  In Section~\ref{sec:LFUV} we focus on the experimental determination of the LECs involved in $K^\pm \to \pi^\pm\ell^+\ell^-$ and $K_L \to \ell^+\ell^-$, and comment on how measurements at NA62 and future kaon experiments may improve the resulting limits on $C_{9,10}^B$.  A similar strategy is adopted in Section~\ref{sec:LFV} to obtain bounds on lepton flavour violation (LFV) in the $B$-meson sector, while concluding remarks are given in Section~\ref{sec:summary}.  For the analysis of LFUV and LFV in other kaon decays not discussed in this article, we refer the reader to~\cite{Crivellin:2016vjc}.

\section{Kaon probes of lepton flavour universality violation}
\label{sec:LFUV}

\subsection{$K^\pm \to \pi^\pm \ell^+\ell^-$}
At low energies, the dominant contribution to $K^+\to\pi^+\ell^+\ell^-$ is due to single virtual-photon exchange. The resulting amplitude involves a vector form factor $V_+(z)$ which up to $O(p^6)$ in the chiral expansion, can be decomposed in the general form~\cite{Cirigliano:2011ny} 
\begin{equation}
  V_+(z) = a_+ + b_+z + V_+^{\pi\pi}(z)\,,  \qquad z = q^2 / m_K^2\,.
  \label{V_dec}
\end{equation}
Here the LECs $a_+$ and $b_+$ parametrise the polynomial part, while the rescattering contribution $V_+^{\pi\pi}$ can be determined from fits to $K\to\pi\pi$ and $K\to\pi\pi\pi$ data.  Chiral symmetry alone does not constrain the values of the LECs,%
  \footnote{Although there are recent attempts to determine $a_+$ and $b_+$ on the lattice~\cite{Christ:2015aha} or via large-$N_c$ methods~\cite{Coluccio-Leskow:2016tsp}.}
so instead, we consider the differential decay rate $d \Gamma / dz \propto |V_+(z)|^2$ as a means to extract $a_+$ and $b_+$ from experiment.  The resulting fit to the decay spectra from all available high-statistics experiments is given in Table~\ref{tab:a+b+}.

\begin{table}[t]
\renewcommand{\arraystretch}{1.3}
\centering
\begin{tabular}{cccr}\toprule 
Channel & $a_+$ & $b_+$ & Reference \\
\hline
$ee$ & $-0.587\pm 0.010$ & $-0.655\pm 0.044$ & E865~\cite{Appel:1999yq}\\
$ee$ & $-0.578\pm 0.016$ & $-0.779\pm 0.066$ & NA48/2~\cite{Batley:2009aa}\\
$\mu\mu$ & $-0.575\pm 0.039$ & $-0.813\pm 0.145$ & NA48/2~\cite{Batley:2011zz}\\\bottomrule
\end{tabular}
\caption{Fitted values of coefficients in the vector form factor~\eqref{V_dec}.}
\label{tab:a+b+}
\end{table}

Now for the crucial point: if lepton flavour universality applies, the coefficients $a_+$ and $b_+$ have to be equal for the $ee$ and $\mu\mu$ channels, which within errors is indeed the case.  Since the SM interactions are lepton flavour universal, deviations from zero in differences like $a_+^{\mu\mu} - a_+^{ee}$ would then be a sign of NP, and the corresponding effect would be necessarily short-distance. 

To convert the allowed range on $a_+^\text{NP}$ into a corresponding range in the Wilson coefficients $C_{7V}^{\ell\ell}$, we make use of the $O(p^2)$ chiral realization of the $SU(3)_L$ current
\begin{equation}
  \bar{s}\gamma^\mu(1-\gamma_5)d \leftrightarrow i F_\pi^2 (U\partial^\mu U^\dagger)_{23}\,, 
  \qquad U = U(\pi,K,\eta)\,, 
  \label{bosonize}
\end{equation}
to obtain 
\begin{equation}
 a_+^\text{NP}=\frac{2\pi\sqrt{2}}{\alpha}V_{ud}V_{us}^*C_{7V}^\text{NP}\,.
 \label{aNP}
\end{equation}
Contributions due to NP in $K^+ \to \pi^+ \ell^+\ell^-$ can then be probed by considering the \emph{difference} between the two channels
\begin{equation}
\label{limit_Kp}
 C_{7V}^{\mu\mu}-C_{7V}^{ee}=\alpha\frac{a_+^{\mu\mu}-a_+^{ee}}{2\pi\sqrt{2}V_{ud}V_{us}^*}\,.
\end{equation}
If the framework of MFV, this can be converted into a constraint on the NP contribution to $C_9^B$:
\begin{equation}
\label{C_charged}
 C_{9}^{B,\mu\mu}-C_{9}^{B,ee}=-\frac{a_+^{\mu\mu}-a_+^{ee}}{\sqrt{2}V_{td}V_{ts}^*}\approx -19\pm 79\,,
\end{equation}
where we have averaged over the two electron experiments listed in Table~\ref{tab:a+b+}.

Evidently, the determination of $a_+^{\mu\mu}-a_+^{ee}$ needs to be improved by an $O(10)$ factor in order to probe the parameter space relevant for the $B$-anomalies, whose explanation involves Wilson coefficients $C_{9,10}^B=O(1)$~\cite{Descotes-Genon:2015uva}. Improvements of this size may be possible at NA62, especially for the experimentally cleaner dimuon mode which currently has the larger uncertainty. 

\subsection{$K_L \to \ell^+\ell^-$}
With an eye towards future kaon experiments involving $K_L$ beams%
  \footnote{LCT thanks Matthew Moulson for informative discussions on the challenges of measuring rare $K_L$ decays with high precision.} 
(e.g.\ a side programme like KLEVER at NA62~\cite{Moulson}), we also consider $K_L\to\ell^+\ell^-$ decays as another potential probe of LFUV.  These decays are complementary to $K^+\to \pi^+\ell^+\ell^-$ since they provide the means to constrain NP effects due to axial-vector interactions.
 
The dominant long-distance contribution is due to $K_L\to\gamma^*\gamma^*\to\ell^+\ell^-$, where the dispersive component of the decay amplitude involves a counterterm $\chi$ that is decomposed in long- and short-distance parts $\chi(\mu) = \chi_{\gamma\gamma}(\mu) + \chi^{}_\text{SD}$. The SM prediction for $\chi^{}_\text{SD}$ is known~\cite{Cirigliano:2011ny}, but $\chi_{\gamma\gamma}$ depends on two LECs whose values are not fixed by chiral symmetry. Nevertheless, we invoke the same argument applied to $K^+\to\pi^+\ell^+\ell^-$, and observe that if lepton flavour universality holds, then the SM values of $\chi$ must be equal in both the $ee$ and $\mu\mu$ channels. Then, using the chiral realization~\eqref{bosonize} of the $SU(3)_L$ current, one obtains an analogous relation to~\eqref{aNP} for the NP Wilson coefficient:
\begin{equation}
\label{Wilson_Kll}
  C_{7A}^\text{NP} = -\frac{\alpha}{F_KG_F V_{ud}V_{us}^* } \bigg(\frac{2\Gamma_{\gamma\gamma}}{\pi \mk^3}\bigg)^{1/2}  \chi^{}_\text{NP}\,,  \qquad \Gamma_{\gamma\gamma} = \Gamma(K_L\to \gamma\gamma)\,.
\end{equation}
The final step is to observe that within the framework of MFV, the difference 
\begin{align}
  C_{7A}^{\mu\mu} - C_{7A}^{ee} &=  -\frac{\alpha}{F_K G_F V_{ud}V_{us}^*} \bigg(\frac{2\Gamma_{\gamma\gamma}}{\pi \mk^3}\bigg)^{1/2} \big( \chi^{\mu\mu} - \chi^{ee} \big) \,,
  \label{chi diff}
\end{align}
is directly related to the Wilson coefficients of the $B$-physics sector:
\begin{align}
 C_{10}^{B,\mu\mu}-C_{10}^{B,ee} =2.6\bigg(\frac{3.5\times 10^{-4}}{V_{td}V_{ts}^*}\bigg)\big( \chi^{\mu\mu} - \chi^{ee} \big)\,.
 \label{eq:C10 diff}
\end{align}

Clearly, the quality of the bounds on $C_{10}^{B,\ell\ell}$ depends on the precision with which $\chi^{\ell\ell}$ can be determined.  The present situation is as follows: $\chi$ can be determined (up to a two-fold ambiguity) from the measured $K_L\to\ell^+\ell^-$ rates, with the resulting values shown in Table~\ref{tab:chi}.%
  \footnote{One subtlety involved in the fit is that the equality $\chi^{\mu\mu} = \chi^{ee}$ is assumed to hold to all orders in the chiral expansion.  In general~\cite{Isidori:2003ts,D'Ambrosio:1997jp}, higher-order corrections involve $m_\ell$-dependent terms that produce a shift $\Delta\chi$ which must be subtracted to identify potential LFUV contributions. At the two-loop level, we have estimated~\cite{Crivellin:2016vjc} the size of the shift to be $\Delta \chi^{\mu\mu} - \Delta \chi^{ee} = -2.8$, in line with explicit calculations based on Canterbury approximants~\cite{Masjuan:2015cjl}.}
Although Solution 2 for the $ee$ channel is easily excluded, the current data are not sufficiently precise to distinguish among the remaining Solutions.  Moreover, Solution 1 for the $ee$ channel carries a large uncertainty which needs to be improved in order to test LFUV in the interesting regions of parameter space.

\begin{table}[t]
\renewcommand{\arraystretch}{1.3}
\centering
\begin{tabular}{ccc}\toprule
Channel & $\chi$ (Solution 1) & $\chi$ (Solution 2) \\
\hline
$ee$ & $5.1^{+15.4}_{-10.3}$ & $-\big(57.5^{+15.4}_{-10.3}\big)$ \\
$\mu\mu$ & $3.75\pm0.20$ & $1.52\pm0.20$ \\\bottomrule
\end{tabular}
\caption{Values of the contact term $\chi(M_\rho)$ extracted from the measured $K_L\to e^+e^-$ and $K_L\to \mu^+\mu^-$ rates.}
\label{tab:chi}
\end{table}

To gain an idea of the improvement in precision required, suppose the uncertainty in $\Gamma(K_L\to\ell^+\ell^-)$ could be reduced by a factor of $10$ and that the central value remained unchanged. In this idealised scenario, Solution 2 for the $\mu\mu$ channel would be strongly disfavored, given that LFUV (if present at all) should manifest itself as a small effect.  Substituting the resulting difference $\chi^{\mu\mu} - \chi^{ee}\sim 1.3\pm 1.3$ into (\ref{eq:C10 diff}) would then yield the bound $C_{10}^{B,\mu\mu}-C_{10}^{B,ee}\approx 3.5\pm 3.5$.  We thus find that the improvement required to obtain competitive bounds on $C_{10}^B$ is of similar magnitude to what we found in the analysis of $C_9^B$ in $K^+\to\pi^+\ell^+\ell^-$.

We note that, contrary to the $K_L$ decay, short-distance contributions to the $K_S\to\ell^+\ell^-$ decay width do not interfere with any long-distance physics. In the expression for the decay width~\cite{Cirigliano:2011ny}
\begin{equation}
  \Gamma(K_S \to \ell^+\ell^-) = \frac{m_K}{8\pi}\beta_\ell \left[ \beta_\ell^2|B|^2 + |C|^2 \right]\,,
\end{equation}
short-distance effects contribute to the second term~\cite{Isidori:2003ts}, which in the notation of~\cite{Cirigliano:2011ny} reads 
\begin{equation}
  \Im C_\mathrm{SD} =  -2\sqrt{2} F_K m_\ell\, \Im\{ V_{ud} V_{us}^* C_{7A} \} = 2\sqrt{2} F_K m_\ell \,y_{7A} \,\Im\{ V_{td} V_{ts}^*\}\,.
\end{equation}
Recently, LHCb have strengthened their 2012 bound~\cite{Aaij:2012rt} on the $K_S\to\mu^+\mu^-$ rate by roughly a factor of two: $\mathrm{BR}(K_S\to\mu^+\mu^-) < 6.9(5.8)\times 10^{-9}$ at the 95\% (90\%) confidence level~\cite{Ramos}.  With improvements expected in the near future, LHCb will start probing the theoretically interesting region $\mathrm{BR}(K_S\to\mu^+\mu^-)\approx 10^{-11}$~\cite{Isidori:2003ts} where meaningful bounds on NP contributions to the coefficient $y_{7A}$ can be derived. 

\section{Lepton flavour violating decays}
\label{sec:LFV}
NA62 is projected to improve existing limits on LFV in the kaon sector by an order of magnitude or more~\cite{Ceccucci:CD2015}.  Here we adopt a similar approach to our analysis of LFUV in Section~\ref{sec:LFUV}, and use MFV to convert limits on the LFV Wilson coefficients $C_{7V,7A}^{\mu e}$ of the kaon sector into bounds on the corresponding $b\to s$ transitions.  The analysis is simplified by the fact that LFV is absent in the SM%
  \footnote{Modulo negligible effects due to neutrino oscillations.}
so the decay amplitudes can be expressed directly in terms of $C_{7V,7A}^{\mu e}$ and quark operators based on the chiral realization~\eqref{bosonize}.

In the context of LFV searches at NA62, the mode of interest to us is $K^+\to\pi^+\mu^\pm e^\mp$, whose branching fraction takes the form
\begin{align}
\label{GammaKpi}
 \Br\big[K^+\to\pi^+\mu^\pm e^\mp\big]&=0.027\big\{|C_{7V}^{\mu e}|^2+|C_{7A}^{\mu e}|^2\big\}\,.
\end{align}

As a point of comparison, the same combination of Wilson coefficients enters in the $K_L\to\mu^\pm e^\mp$ decay, with branching fraction 
\begin{align}
\label{GammaK}
 \Br\big[K_L\to\mu^\pm e^\mp\big]&=2.6\big\{|C_{7V}^{\mu e}|^2+|C_{7A}^{\mu e}|^2\big\}\,.
\end{align}

Based on~\eqref{GammaKpi} and~\eqref{GammaK}, the present experimental limits~\cite{Ceccucci:CD2015} can be used to constrain the term $(|C_{7V}^{\mu e}|^2+|C_{7A}^{\mu e}|^2)^{1/2}$. The resulting limits are given in the first line of Table~\ref{tab:limits}, where the limit from $K_L\to\mu^\pm e^\mp$ decays is an order of magnitude more stringent than the one from $K^+\to\pi^+\mu^\pm e^\mp$. Although the projected limits from NA62~\cite{Ceccucci:CD2015} will also fall short by a factor of $4$, it would be interesting to examine whether removal of the GigaTracker could produce a sufficient increase in statistics to become competitive with the $K_L$ limits.%
  \footnote{The signal for $K^+\to\pi^+\mu^\pm e^\mp$ is three charged tracks, so an accurate measurement of the pion and lepton momenta is not essential. LCT thanks A.~Ceccucci for sharing this suggestion.}

As in the case of LFUV, we use MFV to obtain limits on the $B$-physics coefficients. These are shown in the bottom line of Table~\ref{tab:limits}, where in the case of the $K^+\to\pi^+\mu^\pm e^\mp$ decay, the resulting constraints are slightly better than~\eqref{C_charged}, but of similar magnitude.  The strongest constraint is obtained from the limit on $K_L\to\mu e$.

\begin{table*}[t]
\centering
\begin{tabular}{cccc}\toprule
 &  $K_L\to\mu^\pm e^\mp$ &  $K^+\to\pi^+\mu^\pm e^\mp$   &  $K^+\to\pi^+\mu^\pm e^\mp$ (NA62 projection) \\
\hline
$\big(|C_{7V}^{\mu e}|^2+|C_{7A}^{\mu e}|^2\big)^{1/2}$ & $<1.3\times 10^{-6}$ & $<2.2\times 10^{-5}$ & $<5.1\times 10^{-6}$\\
$\big(|C_{9}^{B,\mu e}|^2+|C_{10}^{B,\mu e}|^2\big)^{1/2}$ & $<0.71$ & $<12$  & $<2.7$\\
\bottomrule
\end{tabular}
\caption{Limits on LFV Wilson coefficients from kaon decays.  The last line shows the corresponding limits in the $B$-system assuming MFV, while the rightmost column refers to the projected limit from NA62~\cite{Ceccucci:CD2015}.}
\label{tab:limits}
\end{table*}

\section{Remarks and future prospects}
\label{sec:summary}
Rare kaon decays offer a potential probe into NP explanations of the flavour anomalies observed by LHCb in semi-leptonic $B$-decays. In the framework of MFV, we have discussed how limits on LFUV and LFV at kaon experiments can be converted into bounds on the Wilson coefficients $C_{9,10}^B$ of the $b\to s$ effective Hamiltonian.  In this respect, we have focused on the $K^+$ decay modes we consider to be of most relevance to the NA62 programme (with $K_L$ modes included as a point of  comparison):

\begin{itemize}
  \item $K^+\to\pi^+\ell^+\ell^-$ and $K_L\to\ell^+\ell^-$ as a means to constrain LFUV due to vector and axial-vector interactions.  Although the LECs involved in these modes are poorly constrained  from theory, they can be determined via precise experimental measurements.  In particular, bounds on the short-distance NP effects can be obtained by considering \emph{differences} between the $ee$ and $\mu\mu$ parameters. For this method to obtain meaningful bounds on $C_{9,10}^B$, we found that order-of-magnitude reductions on the LECs' current uncertainties are required.  At NA62, this will require a measurement of the $K^+\to\pi^+\ell^+\ell^-$ spectrum which significantly reduces the uncertainties of the NA48/2 data.
  \item the LFV modes $K^+\to\pi^+\mu^\pm e^\mp$ and $K_L\to \mu^\pm e^\mp$.  In these decays, the amplitude factorises into short- and long-distance components, so strong bounds can be obtained on the former.  NA62's projected limits on $K^+\to\pi^+\mu^\pm e^\mp$ will fall short of the existing E871 bound on $K_L\to \mu^\pm e^\mp$~\cite{Ambrose:1998us}, so it would be worthwhile considering if the experiment can be adapted to boost the sample of $K^+$ decays involving three charged particles.
\end{itemize}

We note that in order to convert bounds in the kaon sector into those in $B$-physics, we have worked within the framework of MFV.  In general, the potential NP may not satisfy MFV, so from this point of view there are three logical possibilities which can be tested at NA62:
\begin{enumerate}
  \item if the NP explanations for the $B$-meson anomalies are consistent with MFV, then one should see a signal at the sensitivities discussed in Sections~\ref{sec:LFUV} and~\ref{sec:LFV};
  \item the experimental searches at a sensitivity expected from MFV turn out negative.  In this case, one could immediately infer that any NP scenario explaining the $B$-anomalies would require departures from MFV;
  \item a signal is observed at current or slightly improved sensitivity.  As in point 2 above, we could then rule out NP explanations of the $B$-anomalies based on MFV.
\end{enumerate} 

\section*{Acknowledgements}
Financial support by MIUR under the project number 2010YJ2NYW, the DOE (Grant No.\ DE-FG02-00ER41132), and the Swiss National Science Foundation is gratefully acknowledged.

\end{document}